\documentclass{scrartcl}
\usepackage[affil-it]{authblk} 
\usepackage{makeidx}  
\usepackage{amsmath}  
\usepackage{booktabs} 
\usepackage{longtable} 
\usepackage{graphicx} 
\usepackage{subfig} 
\begin{document}
\title{Homonym Detection in Curated Bibliographies: Learning from dblp's Experience}
\subtitle{-- full version --}
\author{%
  Marcel~R.~Ackermann%
  \thanks{email: marcel.r.ackermann{@}dagstuhl.de, ORCID: 0000-0001-7644-2495}
}
\author{%
  Florian~Reitz%
  \thanks{email: florian.reitz{@}dagstuhl.de, ORCID: 0000-0001-6114-3388}
}
\affil{\small
  Schloss Dagstuhl LZI\\
  dblp computer science bibliography\\
  66687 Wadern, Germany
}

\date{\small June 15, 2018}

\maketitle              

\begin{abstract}
\noindent\textbf{Abstract.}
Identifying (and fixing) homonymous and synonymous author profiles is one of the major tasks
of curating personalized bibliographic metadata repositories like the \emph{dblp computer science bibliography}.
In this paper, we present and evaluate a machine learning approach to identify homonymous author
bibliographies using a simple multilayer perceptron setup.
We train our model on a novel gold-standard data set derived from the past years of active, manual curation at the dblp computer science bibliography.
\end{abstract}

\section{Introduction}

The unambiguous attribution of scholarly publications to their authors ranks among the most critical challenges for digital libraries.
Internal user surveys and usage statistics repeatedly show that targeted author queries are predominant in the navigation patterns of those searching for scholarly material.
Additionally, scientific organizations and policy makers often rely on author-based statistics as a basis for critical action.
Universities and research agencies, for example, use publication and citation statistics for their hiring and funding decisions.
In such cases, a correct attribution is essential.

Modern digital libraries are therefore compelled to provide accurate and reliable author disambiguation of their records.
One such database is the \emph{dblp computer science bibliography}, which collects, curates, and provides open bibliographic metadata of scholarly publications in computer science and related fields~\cite{DBLP:journals/pvldb/Ley09}.
The database was established in 1993 by Michael Ley at the University of Trier.
Since 2011, DBLP is a joint service of University of Trier and Schloss Dagstuhl LZI.
As of January 2018, the collection contains metadata for more than 4 million publications, which are listed on more than 2 million author bibliographies.
Every year, about 400,000 new publications are added to the database.

As can be easily seen from those numbers, the enormous growth in scholarly output in recent years has made purely manual curation of author bibliographies impracticable.
Therefore, algorithmic methods for supporting author disambiguation tasks are necessary.
The two most notorious problem categories are: (1)~cases when different persons share the same name (known as the \emph{homonym} problem), and (2)~cases when the name of a particular author is given in several different ways (known as the \emph{synonym} problem).
Furthermore, there are even mixed cases when a person is subject to both the homonym and the synonym problem at the same time.
Due to these problems, incorrect assignments of publications to authors might lead to defective bibliographies.
Hence, we need proper capabilities of detecting such defects.

\subsection{Our contribution}

In this paper, we present and evaluate a machine learning approach to detect homonymous author bibliographies in large bibliographic databases.
To this end, we train a standard multilayer perceptron (e.g., see~\cite[Ch.~2]{PattersonG2017}) to classify an author profile into either of the two classes ``homonym'' or ``non-homonym''.
While the setup of our artificial neural network is pretty standard, we make use of two original components to build our classifier:
\begin{itemize}
\item We use historic log data from the past years of active, manual curation at dblp to build a ``golden'' training and testing data set of more than 24,000 labeled author profiles for the homonym detection task.
\item We define a vectorization scheme that maps inhomogeneously sized and structured author profiles onto numerical vectors of fixed dimension.
The design of these numerical features is based on the practical experience and domain knowledge obtained by the dblp team and uses only a minimal amount of core bibliographic metadata. We also study the impact of the individual feature groups on our classifiers effectiveness.
\end{itemize}
Please note that since our approach has been designed as an effort to improve the dblp computer science bibliography, it (in accordance with dblp's metadata curation philosophy, see Sec.~\ref{sec:dblp-curation}) intends to keep a human curator in the loop and just uncovers defective profiles, instead of trying to algorithmically resolve the defect.
Fully automatic approaches are only briefly discussed in Sec.~\ref{sec:discussion}.

\subsection{Related work}

Author name disambiguation in digital libraries has been the subject of intensive research for decades.
For an overview on different algorithmic approaches see the survey by Ferreira~et~al.~\cite{DBLP:journals/sigmod/FerreiraGL12}.
The vast majority of these approaches tackle author name disambiguation as a batch task by re-clustering all the existing publications at once.
However, in the practice of a curated database like dblp, disambiguation is performed rather incrementally as new metadata is added, and by preserving the curation effort that has been made to the bibliographies in earlier iterations.
Only recently, a number of approaches have been published that consider these practice-driven constraints~\cite{DBLP:journals/jidm/CarvalhoFLG11,DBLP:journals/jidm/EsperidiaoFLGGTA14,DBLP:journals/ir/QianZSYL15,DBLP:journals/jasis/SantanaGLF17,DBLP:conf/dsaa/ZhaoRBR17}.

With the recent advances made in the field of artificial intelligence, a number of (deep) artificial neural network method have also been applied to author name disambiguation problems~\cite{DBLP:conf/aciids/TranHD14,DBLP:conf/ercimdl/Muller17}.
However, those previous approaches focus on learning the semantic similarity of individual publications.
It is still unclear how these approaches can be used to assess the homonymity of a whole author's bibliography, as is required in our scenario.

There exist many data sets derived from dblp that are used to train or evaluate author name disambiguation methods~\cite{DBLP:journals/ir/QianZSYL15,DBLP:conf/sac/HanXZG05,DBLP:conf/jcdl/HanZG05,DBLP:journals/ipm/KangKLJY11,DBLP:conf/jcdl/MomeniM16}.
For a survey and discussion of the individual advantages and disadvantages of these recent data sets see M\"uller~et~al.~\cite{DBLP:journals/scientometrics/MullerRR17}.
All of those data sets are based on a single snapshots of the dblp database, and they concentrate on a narrow (and sometimes biased) selection of publications from dblp.
To the best of our knowledge, there is no data set that considers the evolution of the curated bibliographies in dblp beside the recently published historical corrections test collection of Reitz~\cite{Zen:Reitz2018}, which is the foundation of our contribution (see Sec.~\ref{sec:gold-data-set}).

\section{Learning homonymous author bibliographies}

\subsection{Metadata curation at dblp}\label{sec:dblp-curation}

One of dblp's characteristic features is the assignment of a publication to its individual author (even in the presence of incomplete information and homonymous or synonymous names) and the curation of bibliographies for all authors in computer science.
In order to guarantee a high level of data quality, this assignment is a semi-automated process that keeps the human data curator in the loop and in charge of all decisions.
In detail, for each incoming publication, the mentioned author names are automatically matched against the existing author profiles in dblp using several specialized string similarity functions~\cite{DBLP:conf/f-egc/LeyR06}.
Then, a simple social network analysis (mainly based on the co-author linkage) is performed to rank the potential candidate profiles.
If a matching author profile is found, the authorship record is assigned, but only after the ranked candidate lists have been manually checked by the human data curator.
In addition, missing, incomplete, or erroneous information in either the incoming publication metadata or the matched author profiles is updated, and some further normalization is applied.

In cases that remain unclear even after a curator checked all candidates, a manual in-depth check is performed, often involving external sources.
However, the amount of new publications processed each day makes exhaustive detailed checking impossible, which inevitably leads to some incorrect assignments.
Thus, while the initial checking of assignments ensures an elevated level of data quality, a significant number of defective author profiles still find their way into the database, especially in the case of homonymous and synonymous names.

To further improve the quality of the database, another automated process checks all existing author profile in dblp on a daily basis.
This process is designed to uncover defects that become evident as a result of newly added data or corrected entries.
By analyzing an author profile and its linked coauthor profiles for suspicious patterns, this process can detect probably synonymous profiles~\cite{DBLP:tr/trier/MI06-01}.
For the detection of probably homonymous profiles, no automated process has existed prior to the results presented here, and the dblp team has been largely relying on hints from the community to become aware of such situations~\cite{DBLP:series/lnsn/Reitz013}.
A simple clustering approach has been used to visualize the (in-)coherence of an author profiles coauthor community, yet without providing conclusive information (see Fig.~\ref{fig:coauth-network}).

If a suspicious case of a synonymous or homonymous profile is validated by manual inspection, then the case is corrected by either merging or splitting the author profiles, or by reassigning a selection of publications from one profile to another.
By doing so, in 2017 alone, in~9,731 cases author profiles were merged and a total of~3,254 author profiles have been split, while in~6,213 cases partial profiles have been redistributed.
This curation history of dblp forms a valuable set of ``golden'' training and testing data set for curating author profiles~\cite{Zen:Reitz2018}.

\subsection{A gold-data set for homonym detection}\label{sec:gold-data-set}

We use the historic dblp curation data from the embedded test collection as described by Reitz~\cite[Sec.~3.2]{Zen:Reitz2018} to build a ``golden'' data set for homonym detection.
This collection compares dblp snapshots from different timestamps $t_1 < t_2$ and classifies the manual corrections made to the author bibliographies between $t_1$ and $t_2$.
For this paper, we use the historic data from the dblp log files for the observation interval $[t_1,t_2]$ with $t_1 = \textrm{``2014-01-01''}$ and $t_2 = \textrm{``2018-01-01''}$.
The test collection is available online~\cite{Zen:Reitz2018} under Open Data Commons Attribution License (ODC-By).

Within this test collection, we selected all source profiles from the defect cases of type ``Split'' as our training and testing instances of label class ``homonym''.
That is, these are profiles at timestamp $t_1$ where a human curator at some point later between $t_1$ and $t_2$ decided to split the profile (i.e., the profile has actually been homonymous at timestamp $t_1$).

\enlargethispage{2ex}

Additionally, from all other profiles in the dblp data set at timestamp $t_1$, we selected the profiles which did either (a) contain non-trivial person information like a homepage URL or affiliation information, or (b) at least one of the author's names in dblp ends by a ``magic'' 4-digit number (i.e., the profile has been manually disambiguated~\cite{DBLP:journals/pvldb/Ley09} prior to $t_1$) as instances of label class ``non-homonym''.
This selection makes sense since those profiles had all been checked by a human curator at some point prior to $t_1$, and the profile has not been split in the period between $t_1$ and $t_2$.
While this is not necessarily a proof of non-homonymity, such profiles are generally more reliable than an average, random profile from dblp.

In order to further rule out trivial cases for both labels, we dropped all profiles that at timestamp $t_1$ did list either less than two publications or less than two coauthors.
We ended up with a ``golden'' data set of 2,802 profiles labeled as ``homonym'' and 21,576 profiles labeled as ``non-homonym'' (i.e., a total of 24,378 profiles) from the dblp data set at timestamp $t_1$.
Please be aware that the labels in this data set come with a one-sided error: The cases labeled ``homonym'' are reliable since we have proof of such a correction from the historic dblp test collection. On the other hand, the cases labeled ``non-homonym'' have been constructed heuristically and may not always be correct.

\subsection{Vectorization of author bibliographies}

\begin{figure}[t]
\centering
\subfloat[Sketch of local coauthor community clustering: The central node gets removed, and the remaining connected components form the clusters.]{\label{fig:coauth-network}%
  \begin{minipage}{0.45\textwidth}
    \centering
	 \includegraphics[width=0.7\textwidth]{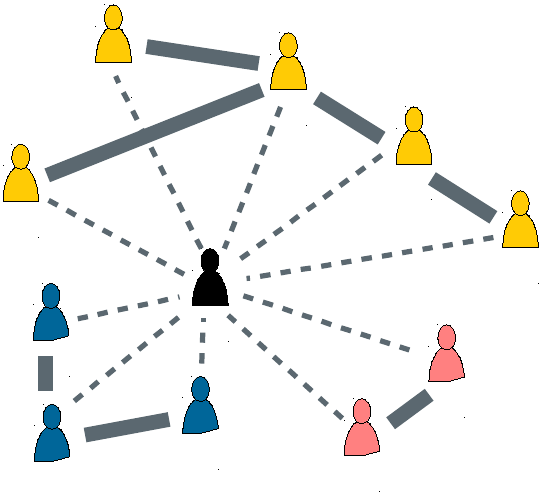}
  \end{minipage}
}
\hspace{2em}
\subfloat[Hyperparameters used to train the title word embedding.]{\label{fig:word2vec-params}%
  \begin{minipage}{0.4\textwidth}
    \centering
    \begin{tabular}{lr} \toprule
    parameter & value \\ \midrule
    embedding dimensions & 150 \\
    minimum word frequency & 2 \\
    context window size & 3 \\
    learning rate & 0.025 \\
    subsampling rate & 1e-5 \\
    iterations & 5 \\
    epochs & 5 \\
    \bottomrule
    \end{tabular}
  \end{minipage}
}
\caption{Auxiliary structures for vectorization.}
\end{figure}

In order to train an artificial neural network using our labeled profiles, we need to represent the non-uniformly sized author profiles at timestamp $t_1$ as numerical vectors of fixed dimension.
To this end, our vectorization makes use of two precomputed auxiliary structures:

\begin{itemize}
\item \emph{local coauthor clusterings:} For each profile, we use a very simple connected component approach to cluster its set of coauthors: First, consider the local (undirected) subgraph of the dblp coauthor network containing only the current person and all direct coauthors as nodes. We call this the local coauthor network. Then, remove the current person and all incident edges from the local coauthor network. The remaining connected components form the coauthor clusters of the current person. See Fig.~\ref{fig:coauth-network} for a small example.
\item \emph{title word embeddings:} We train a vector representation of all title words in the dblp corpus using the \texttt{word2vec} algorithm~\cite{DBLP:conf/nips/MikolovSCCD13}.
In particular, we use the DeepLearning4J~\cite{DL4J} implementation of \texttt{word2vec}, using the skip-gram model and 150 embedding dimensions.
To allow for reproducibility, an overview of the further model hyperparameters\footnote{If our approach is applied to another research domain than computer science, tuning of these hyperparameters might be necessary to improve your results.} is given in Fig.~\ref{fig:word2vec-params}.
In the vectorization below, we use this word embedding model as basis to compute paragraph vectors (also known as \texttt{doc2vec}~\cite{DBLP:conf/icml/LeM14}) of whole publication titles, or even collections of titles.
\end{itemize}

\noindent The design of the feature components of our vectorization is based on the experience and domain knowledge obtained by the dblp team during the years of actively curating the dblp bibliographies.
That is, we identified different features that are implicitly and explicitly taken into consideration whenever a human curator at dblp is assessing the validity of a profile.
In particular, we make use of the following feature groups in our vectors.
A detailed listing of all features is given in Fig.~\ref{fig:features}.
In Section~\ref{sec:evaluation}, we will study the impact of each feature group on the classifier's performance.
\begin{itemize}
\item \emph{group B:} Basic, easy-to-compute facts of the author's profile, i.e., the number of publications, coauthors, and coauthor relations on that profile.
\item \emph{group C:} Features of the local coauthor clustering, like the number of clusters and features of their size distribution. The aim of this feature set is to uncover the incoherence of local coauthor communities, which experience shows to be symptomatic of homonymous profiles.
\item \emph{group T:} Geometric features (in terms of cosine distance) of the embedded paragraph vectors for all publication titles listed on that profile. This feature set aims to uncover inhomogeneous topics of the listed publications, which might be a sign of a homonymous profile.
\item \emph{group V:} Geometric features (in terms of cosine distance) of the embedded paragraph vectors for all venues (i.e. journals and conference series) listed on that profile, where each venue is represented by the complete collection of all titles published in that venue. This feature set also aims to uncover inhomogeneous topics by using the aggregated topical features of its venue as a proxy for the actual publication.
\item \emph{group Y:} Features of the publication years listed on that profile. The aim of this feature group is to uncover profiles that mix up researchers with different years of activity.
\end{itemize}

\begin{figure}[p]
\centering
\begin{tabular}{lllp{10.5cm}}
\toprule
group & dims & type & description \\
\midrule
B & 1 & integer & number of publications of this profile \\
  & 1 & integer & number of coauthors of this profile \\
  & 1 & integer & number of coauthor relations, i.e., the number of edges between coauthors in this author's local coauthor network \\
\midrule
C & 1 & integer & number of local coauthor clusters of this profile \\
  & 5 & integer & sizes of the five largest coauthor clusters, in descending order \\
  & 1 & float & an entropy-inspired measure $h$ of the non-uniformity of the size of the coauthor clusters, in detail: let $n_1,\ldots,n_k$ denote the sizes of the clusters and $N=\sum_{i=1}^k n_i$, then $h = \tfrac{1}{\log k} \sum_{i=1}^k \tfrac{n_i}{N} \log \tfrac{N}{n_i} $ \\
\midrule
T & 1 & float & diameter (in terms of cosine distance) of the embedded paragraph vectors for all publication titles of this profile \\
& 1 & integer & approximate number of centers from the set of the embedded title vectors required, such that all embedded vectors are within cosine distance $0.5$ of at least one center; this number is approximated using Gonzalez' greedy algorithm~\cite{DBLP:journals/tcs/Gonzalez85} \\
& 5 & float & the 5th, 25th, 50th, 75th, and 95th percentile of all pairwise cosine distances between all pairs of embedded title vectors \\
& 5 & float & the 5th, 25th, 50th, 75th, and 95th percentile of the cosine distances from all embedded title vectors to their centroid \\
\midrule
V & 1 & integer & number of venues with publications listed on that profile \\
& 1 & float & diameter (in terms of cosine distance) of the embedded paragraph vectors for all venues listed on that profile, where a venue is represented by the collection of all titles from that venue \\
& 1 & integer & approximate number of centers from the set of the embedded venue vectors required, such that all embedded vectors are within cosine distance $0.5$ of at least one center; this number is approximated using Gonzalez' greedy algorithm~\cite{DBLP:journals/tcs/Gonzalez85} \\
& 5 & float & the 5th, 25th, 50th, 75th, and 95th percentile of all pairwise cosine distances between all pairs of embedded venue vectors \\
& 5 & float & the 5th, 25th, 50th, 75th, and 95th percentile of the cosine distances from all embedded venue vectors to their centroid \\
\midrule
Y & 1 & integer & span of years from earliest to most recent publication of this profile \\
& 1 & integer & number of different years in which an article has been published \\
& 1 & integer & largest gap in years between two publications \\
& 1 & integer & gap in years between the largest and second largest mode in the publications-per-year histogram (using the more recent modes/years if there is a tie), or $0$ if there is only one mode \\
\bottomrule
\end{tabular}
\caption{A detailed listing of all features, sorted by feature group.}\label{fig:features}
\end{figure}

\subsection{Classifier setup}

As classifier we define a standard multilayer perceptron~\cite[Ch.~2]{PattersonG2017} with three hidden layers.
In particular, for each experiment, our classifier has a variable number of input nodes (depending on the concrete selection of feature groups we use in each experiment, see Sec.~\ref{sec:evaluation}), followed by 32 inner nodes in the first hidden layer, 16 nodes in the second, and 8 nodes in the third.
Finally, the output layer consists of two nodes, representing the label classes ``non-homonym'' and ``homonym''.
The activation function used in the hidden layers are rectified linear units (ReLU), while the output layer uses the softmax activation function in order to allow for an interpretation of the output values as a probability distribution.
We use binary cross-entropy as loss function and stochastic gradient descent as optimization algorithm.
L2 regularization is used to fight overfitting.
Further hyperparameters of our classifier are listed in Fig.~\ref{fig:mlp-params}.

\begin{figure}[t]
\centering
\begin{tabular}{lr} \toprule
parameter & value \\ \midrule
1st hidden layer & 32 nodes \\
2nd hidden layer & 16 nodes \\
3rd hidden layer & 8 nodes \\
output layer & 2 nodes \\
hidden layer activation & ReLU \\
hidden layer initialization & ReLU \\
output layer activation & SoftMax \\
output layer initialization & Xavier \\
loss function & binary cross-entropy \\
optimization & stochastic gradient descent \\
learning rate & 0.001 \\
batch size & 32 \\
updater & Adam \\
regularization & L2 ($\alpha=0.001$) \\
iterations & 1 \\
epochs & 40 \\
\bottomrule
\end{tabular}
\caption{Hyperparameters used to train the classifier.}\label{fig:mlp-params}
\end{figure}

\section{Evaluation}\label{sec:evaluation}

\subsection{Implementation}
\enlargethispage{4ex}

We implemented and trained our classifier using the open-source Java library DeepLearning4J~\cite{DL4J}.
While Python-based implementations like Tensorflow or Keras seem to be more commonplace in academic research contexts,
the production environment of dblp and dblp's custom code is mainly based on Java.
Hence, an enterprise-level Java library was the best fit for our live production environment.
All experiments have been conducted on a standard Intel(R) Core(TM) i7-3770 CPU~@~3.40GHz desktop PC, using Java8 and allocating 16\,GB of RAM to the JVM.

Before running our experiments, we randomly split our gold-data profiles into fixed sets of 80\% training and 20\% testing profiles.
Since neural networks work best when data is normalized, we rescaled all profile features to have an empirical mean of $0.0$ and a standard error of $1.0$ on the training data.
For each set of vectorization feature groups we studied, 25 models have been trained independently (i.e., using a different random seed each) on the training profiles and evaluated on the testing profiles.

\subsection{Quality measures}

In information retrieval contexts, the quality of algorithms is often evaluated by measures like precision, recall, and F1-score.
However, in the case of unbalanced label classes, these three measures are known to give misleading scores~\cite{Powers2011}.
Homonym detection is such an unbalanced case.
In fact, in our gold data set we find our labels to be unbalanced with a population prevalence of the homonym defect of~11.4\%.
And there is no reason to believe that this ratio is even representative of a bibliography database as a whole, where experience suggests that the true ratio might be much closer to~1.0\% or even~0.1\%.

Hence, when evaluating homonym classifiers, we propose to rather use other measures like Matthews correlation coefficient (MCC)~\cite{Matthews1975} or the area under the receiver operating characteristic (AUROC)~\cite{DBLP:conf/ijcai/LingHZ03} instead, which are known to yield reliable scores for diagnostic tests even if class labels are severely unbalanced~\cite{Powers2011}.
However, in Fig.~\ref{fig:results}, we still give precision, recall, and F1-score in order to allow for our results to be compared with other studies.

\subsection{Results}

\begin{figure}[t]
\centering
\begin{tabular}{lrccccc} \toprule
features & precision & recall & F1-score & \textbf{MCC} & AUROC \\ \midrule
B & $0.823\pm0.311$ & $0.024\pm0.012$ & $0.047\pm0.022$ & $\textbf{0.130}\pm0.055$ & $0.799\pm0.013$ \\
BC & $0.818\pm0.173$ & $0.057\pm0.016$ & $0.106\pm0.028$ & $\textbf{0.197}\pm0.045$ & $0.842\pm0.005$ \\
BT & $0.542\pm0.177$ & $0.051\pm0.030$ & $0.092\pm0.052$ & $\textbf{0.138}\pm0.060$ & $0.786\pm0.009$ \\
BV & $0.745\pm0.040$ & $0.232\pm0.047$ & $0.350\pm0.068$ & $\textbf{0.372}\pm0.055$ & $0.815\pm0.006$ \\
BY & $0.781\pm0.022$ & $0.153\pm0.014$ & $0.256\pm0.020$ & $\textbf{0.314}\pm0.016$ & $0.820\pm0.004$ \\
BTV & $0.709\pm0.011$ & $0.268\pm0.013$ & $0.389\pm0.015$ & $\textbf{0.393}\pm0.013$ & $0.832\pm0.003$ \\
\textbf{BCTVY} & $\textbf{0.793}\pm0.009$ & $\textbf{0.424}\pm0.011$ & $\textbf{0.552}\pm0.010$ & $\textbf{0.541}\pm0.008$ & $\textbf{0.890}\pm0.002$\\
\bottomrule
\end{tabular}
\caption{The result scores of the classifier on the testing data for the different vectorization feature groups we studied, given as ``mean $\pm$ standard deviation'' of the 25 independently trained classifiers.}\label{fig:results}
\end{figure}

The results of our experiments are summarized in Fig.~\ref{fig:results}.
As can be seen from the MCC scores -- and probably not surprisingly -- our classifier is most effective if all studied feature groups are taken into consideration (i.e., feature set ``BCTVY'').
Note that for this set of features, precision is much higher than recall.
However, this is actually tolerable in our real-world application scenario of unbalanced label classes:
We need to severely limit the number of false-positively diagnosed cases (i.e., we need a high precision) in order to have our classifier output to be practically helpful for a human curator, while at the same time in a big bibliographic database, the ability to manually curate defective profiles is more likely limited by the team size than by the number of diagnosed cases (i.e., recall does not necessarily need to be very high).

One interesting observation that can be made in Fig.~\ref{fig:results} is that the geometric features of the publication titles alone do not seem to be all too helpful (see feature set ``BT'' in Fig.~\ref{fig:results}), while the geometric features of the aggregated titles of the venues seem to be the single most helpful feature group (see feature set ``BV'' in Fig.~\ref{fig:results}).
We conjecture that this is due to mere title strings of individual publications not being expressive and characterful enough in our setting to uncover semantic similarities.
One way to improve feature group T would be to additionally use keywords, abstracts, or even full texts to represent a single publication, provided that such information is available in the database.
However, it should be noted that even in its limited form, feature group T is still able to slightly improve the classifier if combined with feature group V (see feature set ``BTV'' in Fig.~\ref{fig:results}).

\begin{figure}[t]
\centering
\includegraphics[width=1\textwidth]{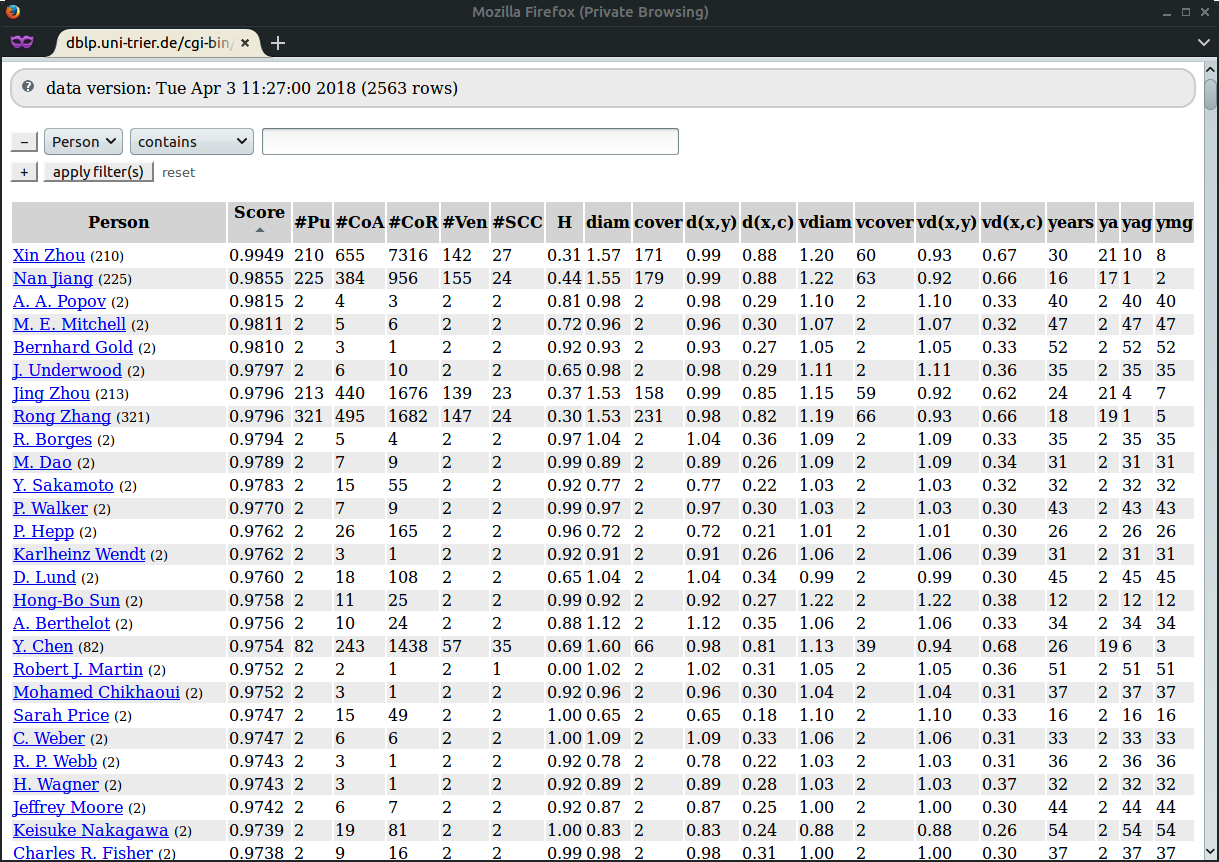}
\caption{Screenshot of a dblp curator's web interface, listing the ranking of probably homonymous profiles in dblp, as inferred by our classifier prototype from the dblp XML dump of April~1,~2018~\cite{DBLP20180401}.}\label{fig:ranking}
\end{figure}

In addition to our experiments, we implemented a first prototype of a continuous homonym detector to be used by the dblp team in order to curate the author profiles of the live dblp database.
To this end, all dblp author profiles are vectorized and assessed by our classifier on a regular basis.
This prototype does not just make use of the binary classification as in our analysis of Fig.~\ref{fig:results}, but rather ranks suspicious profiles according to the probability of label ``homonym'' as inferred by our classifier (i.e., the softmax score of prediction label ``homonym' in the output layer).
The resulting top entries of the ranking are presented to the dblp curators as a web front end in order to easily access, assess, and (if necessary) resolve the suspicious profiles.
A screenshot of the web front end is given in Fig.~\ref{fig:ranking}.
As a small sample from practice, we computed the top 100 ranked profiles from the dblp XML dump of April~1,~2018~\cite{DBLP20180401}, and we checked those profiles manually.
We found that in that practically relevant top list, 74~profiles where correctly uncovered as homonymous profiles, while 12~profiles where false positives, and for 14~profiles the true characteristic could not be determined even after manually researching the case.

\section{Discussion}\label{sec:discussion}

In this paper we presented and evaluated a classifier to detect homonymous author profiles which is motivated by the day-to-day curation work of the dblp computer science bibliography.
Our classifier was made possible by deriving a gold-data set of profiles from the past years of active manual curation.
In order to apply this approach to any other curated bibliography database, a similar extensive history of curation log data is required.
Hence, if such a curation log does not yet exist at your digital library, we strongly encourage you to start collecting such information now in order to enable you to make use of this valuable data set in the future.

However, it should be noted that our vectorization of profiles is based on observations made for the field of computer science, and your mileage may vary if you want to apply it to fields of different characteristics with respect to coauthor communities, choice of publication venues, or frequency of publishing.
Since the actual selection vector features is modular (as demonstrated by our experiments), it should be possible to derive and tune a fitting set of features for another field of study.

Furthermore, our approach is geared towards a scenario where a human curator is taking care of the actual task of fixing the homonymous profile, as is the philosophy employed at the dblp computer science bibliography.
A desired extension of our work is probably a fully automatic approach which also fixes (or at least suggests a solution for) the homonymous profile.
By using the pairwise semantic similarity of publications (e.g.~\cite{DBLP:conf/aciids/TranHD14,DBLP:conf/ercimdl/Muller17}), a clustering of the defective profile might yield such a solution, which is a topic of future research.

\section*{Acknowledgments}\label{sec:acknowledgments}

The research described in this paper was conducted as part of the project ``Scalable Author Disambiguation for Bibliographic Databases'',
funded in part by a grant of the Leibniz Competition, grant no.\ LZI-SAW-2015-2. The authors would like to thank Mark-Christoph M\"uller, Michael Strube, Nicolas Roy, and Adam Bannister for helpful discussion and suggestions.

%
%
\enlargethispage{2ex}
\bibliographystyle{plain}
\bibliography{references}

\begin{thebibliography}{10}

\bibitem{DBLP20180401}
{DBLP}.
\newblock {XML of 2018-04-01}.
\newblock https://dblp.org/xml/release/dblp-2018-04-01.xml.gz.

\bibitem{DBLP:journals/jidm/CarvalhoFLG11}
Ana~Paula de~Carvalho, Anderson~A. Ferreira, Alberto H.~F. Laender, and
  Marcos~Andr{\'{e}} Gon{\c{c}}alves.
\newblock Incremental unsupervised name disambiguation in cleaned digital
  libraries.
\newblock {\em {JIDM}}, 2(3):289--304, 2011.

\bibitem{DBLP:journals/jidm/EsperidiaoFLGGTA14}
Luciano Vilas~Boas Esperidi{\~{a}}o, Anderson~A. Ferreira, Alberto H.~F.
  Laender, Marcos~Andr{\'{e}} Gon{\c{c}}alves, David~Menotti Gomes,
  Andr{\'{e}}a~Iabrudi Tavares, and Guilherme~Tavares de~Assis.
\newblock Reducing fragmentation in incremental author name disambiguation.
\newblock {\em {JIDM}}, 5(3):293--307, 2014.

\bibitem{DBLP:journals/sigmod/FerreiraGL12}
Anderson~A. Ferreira, Marcos~Andr{\'{e}} Gon{\c{c}}alves, and Alberto H.~F.
  Laender.
\newblock A brief survey of automatic methods for author name disambiguation.
\newblock {\em {SIGMOD} Record}, 41(2):15--26, 2012.

\bibitem{DL4J}
Adam Gibson, Chris Nicholson, and Josh Patterson.
\newblock {Eclipse DeepLearning4J v0.9.1}.
\newblock https://deeplearning4j.org.

\bibitem{DBLP:journals/tcs/Gonzalez85}
Teofilo~F. Gonzalez.
\newblock Clustering to minimize the maximum intercluster distance.
\newblock {\em Theor. Comput. Sci.}, 38:293--306, 1985.

\bibitem{DBLP:conf/sac/HanXZG05}
Hui Han, Wei Xu, Hongyuan Zha, and C.~Lee Giles.
\newblock A hierarchical naive bayes mixture model for name disambiguation in
  author citations.
\newblock In {\em {SAC} 2005}, pages 1065--1069. {ACM}, 2005.

\bibitem{DBLP:conf/jcdl/HanZG05}
Hui Han, Hongyuan Zha, and C.~Lee Giles.
\newblock Name disambiguation in author citations using a k-way spectral
  clustering method.
\newblock In {\em {JCDL} 2005}, pages 334--343. {ACM}, 2005.

\bibitem{DBLP:journals/ipm/KangKLJY11}
In{-}Su Kang, Pyung Kim, Seungwoo Lee, Hanmin Jung, and Beom{-}Jong You.
\newblock Construction of a large-scale test set for author disambiguation.
\newblock {\em Inf. Process. Manage.}, 47(3):452--465, 2011.

\bibitem{DBLP:conf/icml/LeM14}
Quoc~V. Le and Tomas Mikolov.
\newblock Distributed representations of sentences and documents.
\newblock In {\em {ICML} 2014}, volume~32 of {\em {JMLR} Proceedings}, pages
  1188--1196. JMLR.org, 2014.

\bibitem{DBLP:journals/pvldb/Ley09}
Michael Ley.
\newblock {DBLP} - some lessons learned.
\newblock {\em {PVLDB}}, 2(2):1493--1500, 2009.

\bibitem{DBLP:conf/f-egc/LeyR06}
Michael Ley and Patrick Reuther.
\newblock Maintaining an online bibliographical database: The problem of data
  quality.
\newblock In {\em EGC 2006}, volume {E-6} of {\em RNTI}, pages 5--10.
  {\`E}d.~C{\'{e}}padu{\`{e}}s, 2006.

\bibitem{DBLP:conf/ijcai/LingHZ03}
Charles~X. Ling, Jin Huang, and Harry Zhang.
\newblock {AUC:} a statistically consistent and more discriminating measure
  than accuracy.
\newblock In {\em IJCAI-03}, pages 519--526. Morgan Kaufmann, 2003.

\bibitem{Matthews1975}
Brian~W. Matthews.
\newblock Comparison of the predicted and observed secondary structure of t4
  phage lysozyme.
\newblock {\em BBA Protein Struct.}, 405(2):442 -- 451, 1975.

\bibitem{DBLP:conf/nips/MikolovSCCD13}
Tomas Mikolov, Ilya Sutskever, Kai Chen, Gregory~S. Corrado, and Jeffrey Dean.
\newblock Distributed representations of words and phrases and their
  compositionality.
\newblock In {\em NIPS 26}, pages 3111--3119, 2013.

\bibitem{DBLP:conf/jcdl/MomeniM16}
Fakhri Momeni and Philipp Mayr.
\newblock Using co-authorship networks for author name disambiguation.
\newblock In {\em {JCDL} 2016}, pages 261--262. {ACM}, 2016.

\bibitem{DBLP:conf/ercimdl/Muller17}
Mark{-}Christoph M{\"{u}}ller.
\newblock Semantic author name disambiguation with word embeddings.
\newblock In {\em {TPDL} 2017}, volume 10450 of {\em LNCS}, pages 300--311.
  Springer, 2017.

\bibitem{DBLP:journals/scientometrics/MullerRR17}
Mark{-}Christoph M{\"{u}}ller, Florian {R}eitz, and Nicolas Roy.
\newblock Data sets for author name disambiguation: an empirical analysis and a
  new resource.
\newblock {\em Scientometrics}, 111(3):1467--1500, 2017.

\bibitem{PattersonG2017}
Josh Patterson and Adam Gibson.
\newblock {\em Deep Learning: A Practitioner's Approach}.
\newblock O'Reilly, 2017.

\bibitem{Powers2011}
David M.~W. Powers.
\newblock Evaluation: From precision, recall and f-measure to roc,
  informedness, markedness \& correlation.
\newblock Technical Report SIE-07-001, Flinders Univ., 2007.

\bibitem{DBLP:journals/ir/QianZSYL15}
Ya{-}nan Qian, Qinghua Zheng, Tetsuya Sakai, Junting Ye, and Jun Liu.
\newblock Dynamic author name disambiguation for growing digital libraries.
\newblock {\em Inf. Retr. J.}, 18(5):379--412, 2015.

\bibitem{Zen:Reitz2018}
Florian Reitz.
\newblock Two test collections for the author name disambiguation problem based
  on {DBLP}, March 2018.
\newblock doi:10.5281/zenodo.1215650.

\bibitem{DBLP:series/lnsn/Reitz013}
Florian {R}eitz and Oliver Hoffmann.
\newblock Learning from the past: An analysis of person name corrections in the
  {DBLP} collection and social network properties of affected entities.
\newblock In {\em The Influence of Technology on Social Network Analysis and
  Mining}, volume~6 of {\em LNSN}, pages 427--453. Springer, 2013.

\bibitem{DBLP:tr/trier/MI06-01}
Patrick Reuther.
\newblock Personal name matching: New test collections and a social network
  based approach.
\newblock Technical Report 06-1, Univ. of Trier, 2006.

\bibitem{DBLP:journals/jasis/SantanaGLF17}
Alan~Filipe Santana, Marcos~Andr{\'{e}} Gon{\c{c}}alves, Alberto H.~F. Laender,
  and Anderson~A. Ferreira.
\newblock Incremental author name disambiguation by exploiting domain-specific
  heuristics.
\newblock {\em {JASIST}}, 68(4):931--945, 2017.

\bibitem{DBLP:conf/aciids/TranHD14}
Hung~Nghiep Tran, Tin Huynh, and Tien Do.
\newblock Author name disambiguation by using deep neural network.
\newblock In {\em {ACIIDS} 2014}, volume 8397 of {\em LNCS}, pages 123--132.
  Springer, 2014.

\bibitem{DBLP:conf/dsaa/ZhaoRBR17}
Zhengqiao Zhao, Jason Rollins, Linge Bai, and Gail Rosen.
\newblock Incremental author name disambiguation for scientific citation data.
\newblock In {\em {DSAA} 2017}, pages 175--183. {IEEE}, 2017.

\end{thebibliography}

\end{document}